\pdfoutput=1
\documentclass[11pt,a4paper]{article}

\usepackage[T1]{fontenc}
\usepackage[utf8]{inputenc}
\usepackage[english]{babel}
\usepackage{mathptmx}                    
\usepackage{microtype}

\usepackage[a4paper,textwidth=17.1cm,textheight=24.0cm,centering]{geometry}

\usepackage{amsmath}
\usepackage{amssymb}
\usepackage{bm}
\usepackage{graphicx}
\usepackage{booktabs}
\usepackage{float}
\usepackage{setspace}
\usepackage[format=plain,justification=justified,singlelinecheck=false,%
            font=small,labelfont=bf,labelsep=space]{caption}
\usepackage[super,sort&compress,comma]{natbib}

\usepackage[hidelinks]{hyperref}

\graphicspath{{./}}

\numberwithin{equation}{section}
\makeatletter
\@namedef{figure*}{\@float{figure}}
\@namedef{endfigure*}{\end@float}
\makeatother

\begin{document}


\hypersetup{pdftitle={AI-assisted analytical theory in soft matter and multiphysics: new results and lessons from diffusiophoresis and bipolar membranes},pdfauthor={Ankur Gupta}}

\begin{center}
{\LARGE\bfseries AI-assisted analytical theory in soft matter and multiphysics: new results and lessons from diffusiophoresis and bipolar membranes\par}
\vspace{1.3em}
{\large Ankur Gupta$^{a,b,c,\ast}$\par}
\vspace{1.0em}
\begin{minipage}{\textwidth}\centering\small
$^{a}$~Department of Chemical and Biological Engineering, University of Colorado Boulder, Boulder, Colorado 80309, USA.\\
$^{b}$~Department of Applied Mathematics, University of Colorado Boulder, Boulder, Colorado 80309, USA.\\
$^{c}$~Materials Science and Engineering Program, University of Colorado Boulder, Boulder, Colorado 80309, USA.\\[3pt]
$^{\ast}$~E-mail: \texttt{ankur.gupta@colorado.edu}
\end{minipage}
\end{center}
\vspace{1.2em}

\begin{center}
\begin{minipage}{0.92\textwidth}
\noindent\textbf{Abstract.}\ We present a framework to advance analytical theory in the area of soft matter and multiphysics using assistance from AI. The framework emphasizes (i) setting ``honesty rules'', (ii) focusing on a few key papers that are used to formulate a problem of interest, (iii) specifying checks a priori, (iv) using only one task per session and keeping a running log of tasks, checks, corrections, and sessions so that a mistake is not propagated through a project, and (v) ensuring all decisions are approved by humans while AI focuses primarily on execution. The framework is applied to two distinct problems to create new knowledge, including (i) diffusiophoresis of a spherical particle in the presence of an arbitrary number of electrolytes at arbitrary Debye lengths, and (ii) analytical description of $I$--$V$ curves, including overlimiting current, in bipolar membranes without including the water dissociation kinetics. Both of these problems are advanced analyses and would each typically require several months to a year's worth of effort, but were solved in a matter of a few days. However, the increased efficiency should be cautiously navigated to avoid cognitive offloading and ensure that AI-assisted results are used to extract human-readable insights instead of just reporting a new mathematical formula. The usage of AI is transparently presented. Finally, we also share brief thoughts on how to integrate AI tools in theoretical research during graduate education.
\end{minipage}
\end{center}
\vspace{1.6em}

\section{Introduction}
\label{sec:intro}

We focus on the potential of AI in shaping future analytical theory in soft
matter and multiphysics. Inspired by prior work in a different field,\cite{Schwartz2026} we
propose a framework to approach analytical theoretical research and use it
to solve two distinct problems: diffusiophoresis of a spherical particle in
a mixture of electrolytes for arbitrary Debye lengths, and analytical
description of $I$--$V$ curves in bipolar membranes, including overlimiting
current. In addition, we share the process and suggestions for researchers
who may be interested in using AI tools in their research.

Some comments are in order before we proceed further. First, we do not
intend to suggest that AI tools should replace human researchers in the
entire process.
On the contrary, we believe that, with careful usage, AI can be used to
improve the process; the framework provided below emphasizes that problem
setup and decision-making should remain under the control of researchers,
and AI should be used primarily for coding and lengthy mathematical
calculations. We understand that modern AI tools are sufficiently advanced
that multiple agents working together can probably solve graduate-level
problems through a single prompt. However, we do not think that such an
approach is necessarily desirable since the AI tools can overwhelm a
researcher, especially an inexperienced one, with their dense and detailed
output. As such, the increased speed comes at the cost of losing the ability
to thoroughly check AI's work and/or to make it accessible to other
researchers in the field and beyond.

Second, while we could solve each of the problems detailed below in a few
days using
AI, we do not think it is realistic for a new researcher in the field to do
so. This was possible because we had thought about these problems for a while
and had a relatively clear attack plan before the project began. Therefore,
in our opinion, a new researcher should first spend time working on
improving their knowledge and developing sufficient skill sets before using
AI. This ensures that they avoid cognitive offloading and can guide the AI
tools effectively and also check their work.

Finally, we are not experts in using AI; we acknowledge that the method we
used here may not be the most effective, and a reader is encouraged to
explore other methods. Having said that, we believe the basic principles of
the framework, i.e., use AI to accelerate the mechanical parts of research
but retain the decision-making, should be easily adaptable to other methods.

The article is organized as follows. Section~\ref{sec:framework} proposes
the framework and explains why this method is proposed, and also clarifies
that some variations may be required for different problems.
Section~\ref{sec:diffusiophoresis} utilizes the framework for a new problem
in diffusiophoresis, and Section~\ref{sec:bpm} repeats the same for a
problem in bipolar membranes. We note that, given that two distinct topics
are solved within a single manuscript, the introduction and method details for both
problems are brief. This decision was made to improve the readability of the
article. Finally, in Section~\ref{sec:education}, we provide some brief
thoughts on graduate training, Section~\ref{sec:conclusion} concludes, and
Section~\ref{sec:si} lists the files provided with the article.

\section{The framework}
\label{sec:framework}

\begin{figure*}[t]
  \centering
  \includegraphics[width=0.8\textwidth]{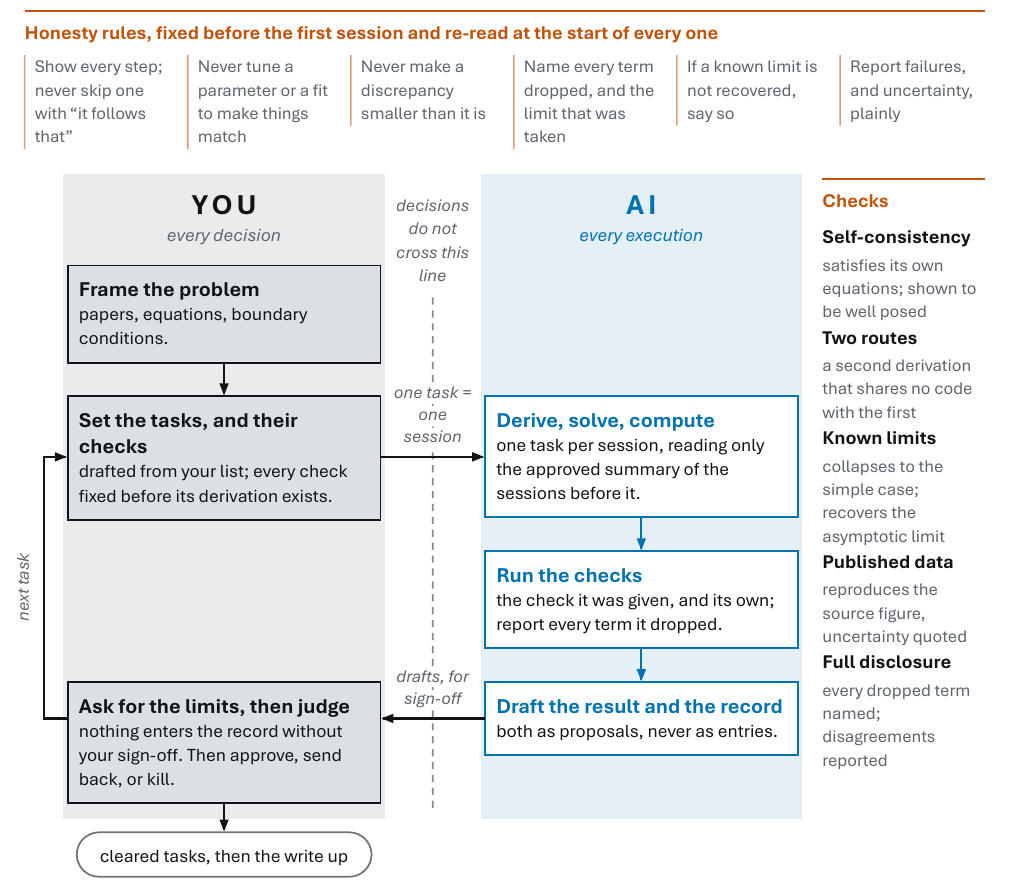}
  \caption{The proposed framework. The honesty rules are fixed before the first session and apply globally. The work is divided into two categories, and decisions never cross the dashed line between them. You own the left category: you frame the problem from the key papers, you set the tasks and the checks each one has to pass, and when a task comes back, you ask for its limits before you approve it, send it back, or kill it. AI owns the right category: it derives, solves, and computes one task per session, reading only the approved summary of the sessions before it, then runs the checks and drafts both the result and the record as proposals rather than entries. The loop repeats task by task until the cleared tasks are ready for the write-up. The five checks on the right are representative ones, applied to relevant tasks. Some dynamic updates to the process may be required depending on the project.}
  \label{fig:framework}
\end{figure*}

The basic framework is presented in Fig.~\ref{fig:framework}. One of the
challenges with AI assistance in theoretical research is that it can arrive
at a plausible result quickly and opaquely. We experienced this firsthand in
our prior work, where a seemingly accurate answer was presented without a
transparent calculation.\cite{GangulyGupta2026} We had to spend significant
effort parsing through the calculations before concluding that there was a
subtle error and the final answer was, in fact, incorrect. This is consistent
with other reports in the literature. Therefore, we propose setting honesty
rules for your full project such as: (i) do not fit a parameter to make
results match, (ii) do not skip derivation steps, (iii) if something failed,
state it plainly and explain why it failed, (iv) do not make discrepancies
smaller artificially, (v) if a known limit is not recovered, say so, and do
not steer the derivation toward a known answer, and (vi) say plainly when you
are unsure. This is not an exhaustive list, and you may have to think through
other sources carefully. However, having these stated upfront minimizes the
chances of error propagation.

Many theoretical advances in soft matter and multiphysics build on progress
in the field. As such, most researchers spend time focusing on a few key
papers, which they then extend to conduct their own analyses. Identifying
these papers needs training and practice, and we suggest spending time to
hone this part of the research. AI may assist in learning and initial
training on the subject itself, but you must understand the papers first
before you extend them. Once you feel comfortable with the literature and
have a clear question/analysis you want to pursue, we suggest giving AI tools
access to these papers only. We then suggest listing out your idea in
explicit detail (see Appendix~\ref{app:goals}), including detailed steps to
follow,
while also planning in advance how to check the work. This helps
systematically lay out the project for an AI tool. Without a guided approach,
AI will set up its own approach and verifications, and they may look very
different from your envisioned plan of attack.

Next, before executing your plan, we suggest breaking it down into a series
of tasks and checks. The idea is to find a balance in giving AI tools a plan
of action, but slowing down the autonomy so that you are following the
calculation as it proceeds. In the project outlined below, we broke down the
project into a few tasks that we felt would be good checkpoints. We used AI
tools to further refine them and break them down into subtasks, but this
further breakdown does not appear to be necessary. We also note that some of
the tasks and checks may need to be updated as the project goes along.
Moreover, we ensured that there is a running log of tasks, checks,
corrections, project summary, and sessions that is generated after each
session; each session is only focused on a specified task so as to avoid
context overload. Finally, each session is given the direction to only read
tasks, checks, and summary files, and not to access other sessions' details.
This helps increase redundancy in calculation and avoid error propagation.

Finally, at the end of each session, AI is used to update the files, and then
you approve them. AI language may sometimes be hard to digest, and the
mathematical details may be too dense to make sense of. In such a scenario,
it is preferred to add prompts to ensure that you are following along. Once
the files are approved, a new session is started for a new task, and the
cycle repeats until the project goals are met. We note that the analysis in
this manuscript was conducted using the Claude Code Opus 5.0 model.


\section{Diffusiophoresis in a mixture of electrolytes}
\label{sec:diffusiophoresis}

\subsection{Background}

Diffusiophoresis is the motion of a colloidal particle due to a concentration gradient of a dissolved solute. Derjaguin and co-workers first reported the phenomenon several decades ago\cite{Derjaguin1947}. Anderson, Prieve and co-workers pursued diffusiophoresis in-depth\cite{AndersonPrieve1984,Anderson1989} and calculated diffusiophoretic velocities in neutral solutes\cite{Anderson1982} and electrolytes\cite{Prieve1984}. Diffusiophoresis has been used to measure zeta potentials\cite{ShinAdvMater2017}, to filter water without a membrane using dissolved carbon dioxide\cite{ShinNatComm2017,ShimStone2020}, to drive colloids into and out of dead-end pores\cite{Shin2016,Kar2015}, to remove particles from porous materials during rinsing\cite{Shin2018}, and to organise colloids into two-dimensional bands\cite{RajShieldsGupta2023}. In biology, a diffusiophoretic mechanism has been proposed for ATP-driven transport that requires no motor proteins\cite{Ramm2021}, and the same mechanism has since been shown to move microtubules along tubulin, RanGTP and salt gradients with no motor present\cite{ShimRammStone2024}. Diffusiophoresis can sharpen Turing patterns\cite{AlessioGupta2023} and, once the colloids are given a finite size and treated as hard spheres, it produces the textured and multiscale patterns that natural systems actually display\cite{MirfendereskiGupta2026}. Several reviews now exist on this topic\cite{Shim2022ChemRev,AultShin2025,Velegol2016,GangulyAlessioGupta2023}.

This paper focuses on diffusiophoresis of a charged sphere in the presence of a concentration gradient of multiple ions. Before we dwell deeper into the mathematical setup, we would like to introduce the physical mechanism briefly. The diffusiophoretic velocity of a charged sphere separates into two contributions that are physically distinct and are treated independently\cite{Prieve1984,KehWei2000}. The first is chemiphoresis, which arises from the nonuniform adsorption of counterions and the depletion of co-ions around the particle: the salt gradient makes the double layer thicker on one side of the particle than on the other, and the resulting osmotic imbalance drives a slip along the surface. The second is electrophoresis. When the cation and the anion of a salt diffuse at different rates, the faster ion runs ahead of the slower one, and the charge separation that this produces sets up a macroscopic electric field which then acts on the charged particle just as an externally applied field would. The two contributions enter at different orders in the surface potential. Writing $\tilde\zeta=e\zeta/k_BT$, where $\zeta$ is the surface potential, $e$ the elementary charge and $k_BT$ the thermal energy, the $O(\tilde\zeta)$ term is electrophoretic and carries the entire dependence on the ion diffusivities through $\beta=(D_+-D_-)/(D_++D_-)$, where $D_+$ and $D_-$ are the cation and anion diffusivities, while the $O(\tilde\zeta^2)$ term is chemiphoretic and does not involve ionic diffusivities.

\subsection{Literature}

The most common result assumes that the double layer is thin compared with the particle, though results with finite Debye length were presented even in the original analysis\cite{Prieve1984}. Keh and Wei removed that restriction\cite{KehWei2000}, obtaining the mobility at low potential but arbitrary double-layer thickness, so that both contributions above acquire a dependence on $\kappa=a/\lambda_D$, where $a$ is the particle radius and $\lambda_D$ the Debye length set by the total ionic strength. Keeping $\kappa$ finite matters whenever the particle is small or the solution dilute. For a particle of radius $50$~nm in a millimolar monovalent salt, $\lambda_D\approx10$~nm and $\kappa\approx5$, and the thin-layer expressions are then in error by tens of per cent.

Both of these treatments consider a single binary electrolyte, while many systems of interest contain several\cite{ShiSquires2016,ColemanGupta2025,FloreaWyss2014}. The ions of different salts do not diffuse independently, because they share one electric field, and prior work has shown that the diffusion of multiple electrolytes cannot be decomposed into separate binary problems even in a one-dimensional pore\cite{Gupta2019SoftMatter}. Multi-ion diffusiophoresis was introduced by Chiang and Velegol\cite{ChiangVelegol2014} for small potentials, and the consequences for particle motion have since been worked out for mixtures of valence-asymmetric electrolytes\cite{GuptaRallabandiStone2019}, for multivalent electrolytes\cite{Wilson2020}, for concentrated solutions\cite{GuptaShimStone2020,Prieve2019}, and for two-dimensional geometries in which diffusiophoresis and diffusioosmosis act in tandem\cite{Alessio2021,Alessio2022}. All of this work assumes the double layer to be thin.

The two generalisations have therefore been pursued along separate lines: arbitrary Debye length for one binary salt, and arbitrary mixtures in the thin-layer limit. Their intersection, i.e., an arbitrary number of ionic species at arbitrary Debye length, has remained open. This is the focus of this project. The route we take is the mobility framework of Ganguly, Roychowdhury and Gupta\cite{GangulyRoychowdhuryGupta2024}, which provides the phoretic velocity of a particle as a single integral over the equilibrium fields for an arbitrary interaction potential.

When we first set out to solve this problem using AI assistance, we had not anticipated the result we eventually obtained. The mechanism of electrophoresis and chemiophoresis described above, in which the diffusivities impact the electrophoretic term, and the chemiphoretic term does not depend on diffusivity, turns out to be a property of the thin-layer limit. At finite $\kappa$, the diffusivities appear in chemiphoresis as well. However, the diffusivity dependence vanishes identically as $\kappa\to\infty$, consistent with prior reports and our understanding of the phenomena.

\begin{figure}[t]
  \centering
  \includegraphics{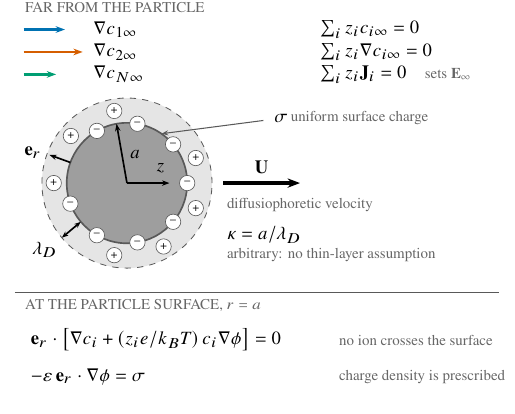}
  \caption{The diffusiophoresis problem. A rigid sphere of radius $a$, carrying a uniform surface charge density $\sigma$, is in an unbounded electrolyte with $N$ dissolved ionic species of valence $z_i$ and diffusivity $D_i$. Each species carries its own imposed uniform gradient $\nabla c_{i\infty}$ far from the particle, constrained only by electroneutrality of the ambient solution and of its gradient. The diffuse charge cloud has thickness $\lambda_D$ set by the total ionic strength, and no assumption is made about $\kappa=a/\lambda_D$. The surface is impermeable to every ion. The particle translates at $\mathbf{U}$, given by eqn~\eqref{eq:dpanswer}.}
  \label{fig:dpschematic}
\end{figure}

\subsection{Problem setup and how AI tools were used}

A schematic of the problem is provided in Fig.~\ref{fig:dpschematic}. A charged spherical particle of radius $a$ is in a Newtonian fluid of viscosity $\mu$ and electrical permittivity $\varepsilon$. The fluid consists of $N$ dissolved ionic species of concentration $c_i$, valence $z_i$, and diffusivity $D_i$. Far from the particle, each species carries its own independently imposed uniform gradient $\nabla c_{i\infty}$, subject to electroneutrality of the ambient solution and of its gradient, $\sum_iz_ic_{i\infty}=0$ and $\sum_iz_i\nabla c_{i\infty}=0$. The surface of the particle is impermeable to every ion and carries a uniform charge density $\sigma$. No assumption is made about the double-layer thickness $\lambda_D$, so $\kappa = a/\lambda_D$ is arbitrary throughout. The objective was to calculate the induced velocity $U$ due to diffusiophoresis in the Debye--H\"uckel limit, or small surface charge. We also assume the weak-field approximation, i.e., the electric fields due to concentration gradients and the surface can be superposed.

\begin{figure*}[t]
  \centering
  \includegraphics[width=0.65\textwidth]{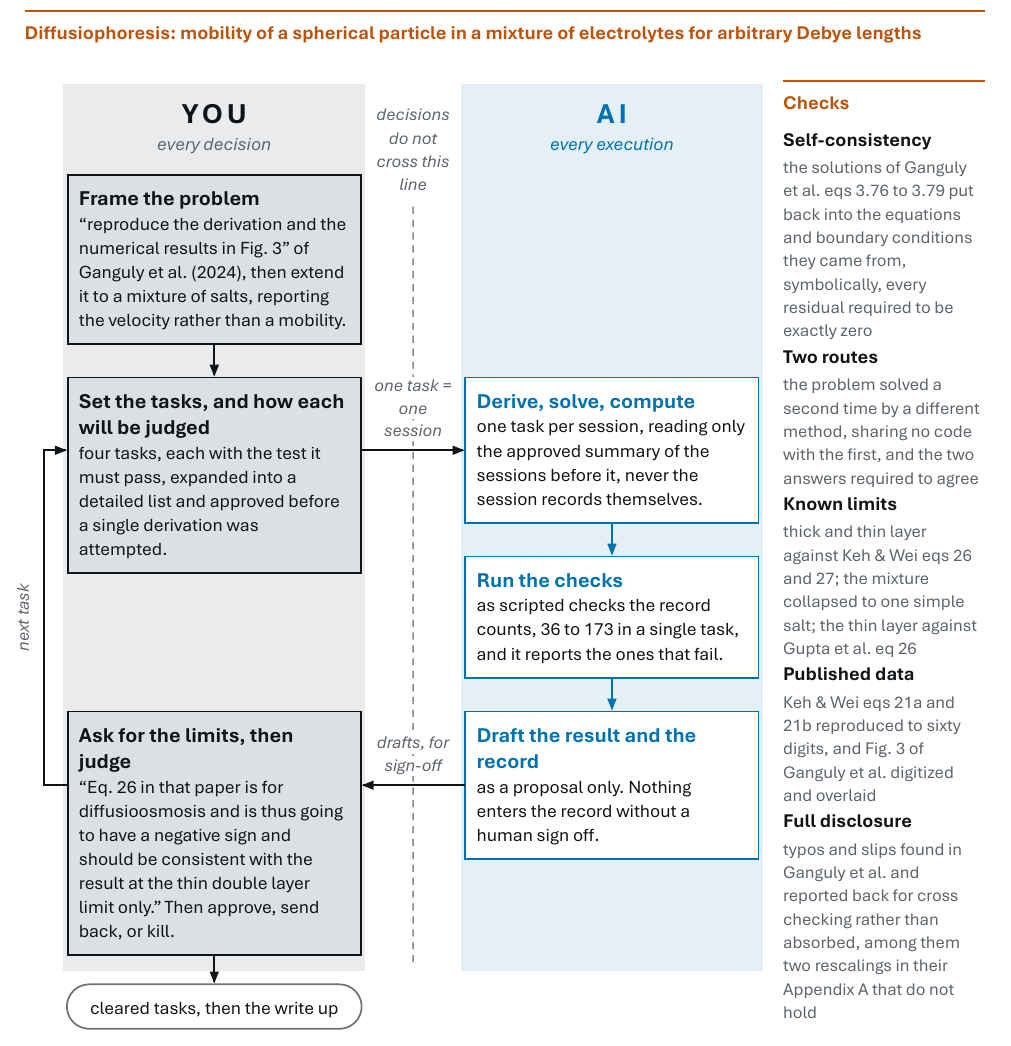}
  \caption{The framework of Fig.~\ref{fig:framework} as it was implemented. Appendix~\ref{app:goals} reproduces the goals and tasks file that set them.}
  \label{fig:dpcase}
\end{figure*}

The plan was fixed before any calculation began; see Fig.~\ref{fig:dpcase} for the layout of how the framework was used. We provided four numbered tasks, reproduced in full in Appendix~\ref{app:goals} and with representative session prompts in Appendix~\ref{app:prompts}, each carrying the checks it had to pass and the deliverable it had to produce. We note that the four tasks written at the start became twenty-one sessions and twenty-two task rows; the plan was refined with help from AI so that the tasks were broken down further. This is not necessary, but it helped us keep track of the progress slightly better. To give some examples of what we input, the first task asked the model to read one section of ref.~\citenum{GangulyRoychowdhuryGupta2024} and to \textit{`reproduce the derivation and the numerical results in Fig. 3'}, with the check being a one-to-one match against the published figure. The fourth task asked for the general result that will be derived to match the thin double layer results of Gupta et al.\cite{GuptaRallabandiStone2019}. We did not make any significant changes in the plan.

The route that the AI should take during calculation was specified in advance as well. The point of sending the model to ref.~\citenum{GangulyRoychowdhuryGupta2024} first was to fix how the velocity would be computed. We also provided some intuition that we had beforehand, while also clarifying the scope of the calculation. For instance, the second task asked whether the problem could be closed on a small number of variables rather than on $N$ separate species equations, in these words: \textit{`Confirm if it is possible to solve the problem only by solving ``salt'', ``charge'', and ``ionic strength'', and maybe one or two more variables.'} In the scope, the third task fixed the order of the expansion at $\tilde\zeta^3$ and asked for the equations and boundary conditions to be collected and checked for well-posedness before anything was solved.

\subsection{Governing equations and the solution procedure}

At small P\'eclet number, $Pe=U_0a/D_{\rm ref}\ll1$ with $U_0$ the particle velocity scale and $D_{\rm ref}$ any reference diffusivity, advection drops out of the ion fluxes and the steady transport and Poisson equations read
\begin{equation}
\nabla\cdot\!\left[\nabla c_i+\frac{z_ie}{k_BT}c_i\nabla\phi\right]=0,
\qquad
-\varepsilon\nabla^2\phi=e\sum_{i}z_ic_i ,
\label{eq:dpgov}
\end{equation}
where $\phi$ is the electric potential. The diffusivity has been divided out of the left-hand equation species by species: the ion distributions do not depend on the diffusivities, which enter only through the far field. At the surface, impermeability and the surface charge condition give
\begin{equation}
\mathbf{e}_r\cdot\!\left[\nabla c_i+\frac{z_ie}{k_BT}c_i\nabla\phi\right]_{r=a}\!\!=0,
\qquad
-\varepsilon\,\mathbf{e}_r\cdot\nabla\phi|_{r=a}=\sigma ,
\label{eq:dpbc}
\end{equation}
with $\mathbf{e}_r$ the outward radial unit vector. The far field is the imposed state of eqn~\eqref{eq:dpgov} plus one further condition: no net current flows at infinity, $\sum_iz_i\mathbf{J}_i|_{r\to\infty}=0$, where $\mathbf{J}_i$ is the flux of species $i$. This sets the conditions
\begin{equation}
\mathbf{E}_\infty=\frac{k_BT}{e}\,
\frac{\sum_iz_iD_i\nabla c_{i\infty}}{\sum_iz_i^2D_ic_{i\infty}} .
\label{eq:dpEinf}
\end{equation}
Equation~\eqref{eq:dpEinf} is the only place the diffusivities enter the problem. Equations~\eqref{eq:dpgov}--\eqref{eq:dpEinf} form a well-posed problem and can be used to calculate ionic concentrations and electric potential.

To calculate the diffusiophoretic velocity $U$, we use the result of Ganguly, Roychowdhury and Gupta\cite{GangulyRoychowdhuryGupta2024}, who apply the reciprocal theorem to a sphere and turn the particle velocity into an integral of the body force over the whole fluid volume $V$,
\begin{equation}
\mathbf{U}=\frac{1}{6\pi\mu a}\int_V\left[
\left(\frac{3a}{2r}-\frac{a^3}{2r^3}-1\right)\mathbf{b}_\perp
+\left(\frac{3a}{4r}+\frac{a^3}{4r^3}-1\right)\mathbf{b}_\parallel\right]dV ,
\label{eq:dpRT}
\end{equation}
where $\mathbf{b}_\perp$ and $\mathbf{b}_\parallel$ are the components of $\mathbf{b}$ normal and parallel to the particle surface. The components of $\mathbf{b}$ include both the electrophoretic force and the excess osmotic force\cite{GangulyRoychowdhuryGupta2024}. Mathematically, $\mathbf{b}$ is given as
\begin{equation}
\mathbf{b}=-e\Big(\sum_iz_ic_i\Big)\nabla\phi
-k_BT\,\nabla\Big(\sum_ic_i-\sum_ic_i^{\rm imp}\Big) ,
\label{eq:dpbody}
\end{equation}
where $c_i^{\rm imp}$ is the imposed far-field state that $c_i$ approaches. The subtraction in the osmotic term is to ensure we only focus on the excess osmotic force.

Therefore, the first step is to calculate concentrations and potential, and then use \eqref{eq:dpRT} to calculate the velocity. To make analytical progress, we perform perturbation analysis by expanding in two small parameters. The first small parameter is the scaled surface potential, $\tilde\zeta=e\zeta/k_BT$. The second is the amplitude of the imposed gradient, $\alpha$, defined by $\mathbf{e}_z\cdot\nabla c_{i\infty}=\alpha n_2\gamma_i/a$, where $\gamma_i$ is the share of the gradient carried by species $i$ and $n_2\equiv\sum_iz_i^2c_{i\infty}$ is twice the ionic strength. Both are taken to be small, i.e., $\tilde\zeta=\frac{e\zeta}{k_BT}\ll1$ and $\alpha\ll1$. We performed the expansion until $O(\alpha)$, which is what makes the velocity linear in the imposed gradients, but we go up to $O(\tilde\zeta^3)$, one order beyond what the published finite-$\kappa$ results reach.

\begin{figure*}[t]
  \centering
  \includegraphics{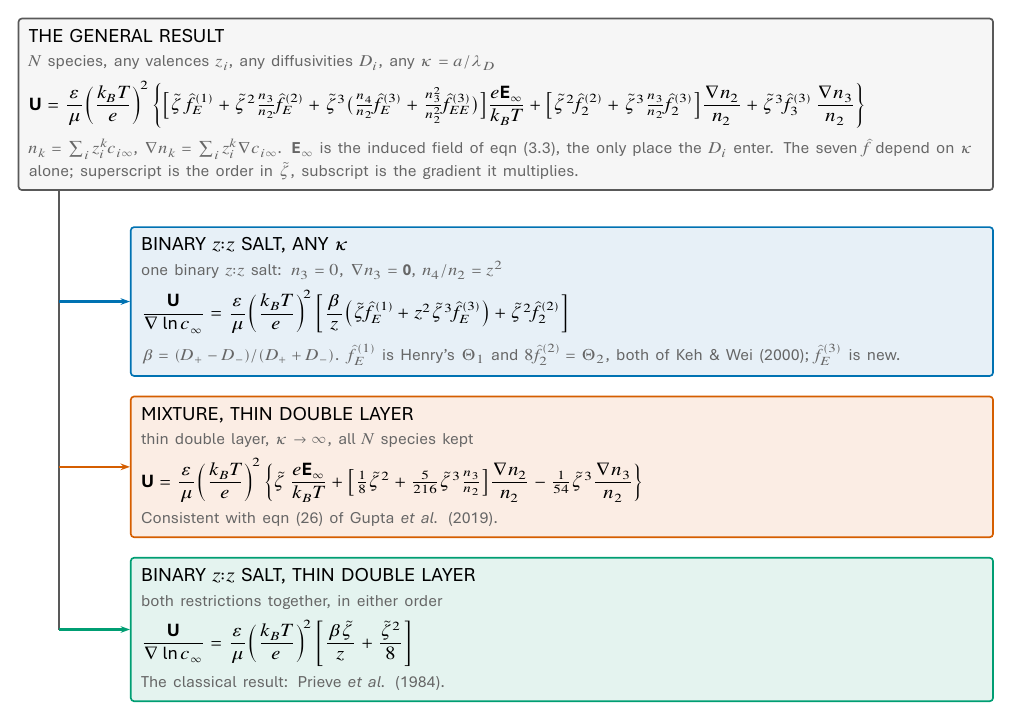}
  \caption{The general result and how it relates to the known results. The top box is eqn~\eqref{eq:dpanswer}, for $N$ ionic species at any $\kappa$. Each row below it is one restriction of that equation: one binary $z{:}z$ salt at arbitrary $\kappa$; a mixture with a thin double layer, which returns eqn~(26) of ref.~\citenum{GuptaRallabandiStone2019}; and both restrictions together, which is the classical result. The last row is reached from either of the two above it, and those two take their limits in opposite orders and agree. Here $\beta=(D_+-D_-)/(D_++D_-)$, $c_\infty$ is the salt concentration of the binary case, and $\mathbf{E}_\infty$ is the induced field of eqn~\eqref{eq:dpEinf}.}
  \label{fig:dpreductions}
\end{figure*}

\begin{figure}[t]
  \centering
  \includegraphics{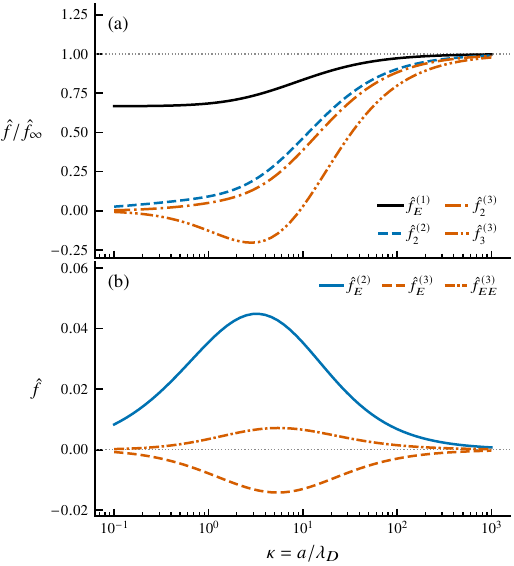}
  \caption{The seven universal functions of $\kappa$ in eqn~\eqref{eq:dpanswer}. Write $\hat f_\infty$ for the value a function takes when the double layer is thin, $\kappa\to\infty$. (a) The four whose $\hat f_\infty$ is nonzero, each divided by its own $\hat f_\infty$ so that all four tend to unity: the limits are $1$ for $\hat f^{(1)}_{E}$, $1/8$ for $\hat f^{(2)}_{2}$, $5/216$ for $\hat f^{(3)}_{2}$ and $-1/54$ for $\hat f^{(3)}_{3}$. (b) The three whose $\hat f_\infty$ is exactly zero; these are the three that multiply $\mathbf{E}_\infty$ beyond leading order, and their vanishing is why no thin-layer treatment contains them.}
  \label{fig:dpuniversal}
\end{figure}

\subsection{Key result}

Rather than take the reader through the hierarchy, we give the key result and then show that it reduces, by itself, to everything that is already known. Fig.~\ref{fig:dpreductions} collects the result and its limits in one place; this subsection states them.

We define valence-weighted moments of the far field,
\begin{equation}
n_k=\sum_i z_i^{k}c_{i\infty} ,
\qquad
\nabla n_k=\sum_i z_i^{k}\nabla c_{i\infty} ,
\label{eq:dpmoments}
\end{equation}
of which $n_1=0$ and $\nabla n_1=\mathbf{0}$ are the two electroneutrality conditions and $n_2$ is twice the ionic strength. In terms of these, the velocity is
\begin{equation}
\begin{aligned}
\mathbf{U}=\frac{\varepsilon}{\mu}\left(\frac{k_BT}{e}\right)^{\!2}\bigg\{
&\Big[\tilde\zeta\hat f^{(1)}_{E}
+\tilde\zeta^{2}\tfrac{n_3}{n_2}\hat f^{(2)}_{E}
+\tilde\zeta^{3}\Big(\tfrac{n_4}{n_2}\hat f^{(3)}_{E}
+\tfrac{n_3^{2}}{n_2^{2}}\hat f^{(3)}_{EE}\Big)\Big]\frac{e\mathbf{E}_\infty}{k_BT}\\
&+\Big[\tilde\zeta^{2}\hat f^{(2)}_{2}
+\tilde\zeta^{3}\tfrac{n_3}{n_2}\hat f^{(3)}_{2}\Big]\frac{\nabla n_2}{n_2}
+\tilde\zeta^{3}\hat f^{(3)}_{3}\,\frac{\nabla n_3}{n_2}\bigg\} ,
\end{aligned}
\label{eq:dpanswer}
\end{equation}
where $\mathbf{E}_\infty$ is the induced field of eqn~\eqref{eq:dpEinf}. On each coefficient, the superscript is its order in $\tilde\zeta$, and the subscript is the gradient it multiplies: $E$ for $\mathbf{E}_\infty$, $2$ for $\nabla n_2$ and $3$ for $\nabla n_3$. The $\tilde\zeta^3$ term multiplying $\mathbf{E}_\infty$ needs two coefficients, one riding $n_4/n_2$ and one riding $n_3^2/n_2^2$, and these are distinguished as $\hat f^{(3)}_{E}$ and $\hat f^{(3)}_{EE}$. Fig.~\ref{fig:dpreductions} summarizes the results in a figure and the various limits it relaxes to. The seven functions are plotted in Fig.~\ref{fig:dpuniversal}.

We focus on two features of eqn~\eqref{eq:dpanswer}. First, the double-layer physics and the physical chemistry separate cleanly: the seven $\hat f$ depend on $\kappa$ and on nothing else, so they can be computed once and used for any combination of electrolytes, while $n_3/n_2$, $n_4/n_2$, $\nabla n_2$ and $\nabla n_3$ depend only on which ions are present. Second, the diffusivities appear nowhere except inside $\mathbf{E}_\infty$, so the bracket multiplying it is the entire diffusivity-dependent response of the particle.

We now focus on the seven functions. Two of the seven are classical. $\hat f^{(1)}_{E}$ is Henry's function $\Theta_1$, the $O(\tilde\zeta)$ electrophoretic coefficient, and $8\hat f^{(2)}_{2}$ is Keh and Wei's $\Theta_2$, the $O(\tilde\zeta^2)$ chemiphoretic coefficient; see Fig.~\ref{fig:dpreductions}. The remaining five exist only for a mixture. We now show how the result is consistent with the known results.

Setting one binary monovalent salt forces $n_3=0$, $\nabla n_3=\mathbf{0}$ and $n_4=n_2$, which switches off $\hat f^{(2)}_{E}$, $\hat f^{(3)}_{2}$, $\hat f^{(3)}_{3}$ and $\hat f^{(3)}_{EE}$. This returns exactly the two classical functions plus one survivor at third order, $\hat f^{(3)}_{E}$. This was a result we did not anticipate. The classical statement that the diffusivities only impact the electrophoretic term is true as $\kappa\to\infty$, but not true at finite $\kappa$.

The seven functions have an asymptotic expansion in inverse powers of $\kappa$. Keeping the leading term of each collapses eqn~\eqref{eq:dpanswer} to the thin-double-layer result shown in Fig.~\ref{fig:dpreductions}, which returns eqn~(26) of ref.~\citenum{GuptaRallabandiStone2019}.

\subsection{Experience with AI tools}

The AI tools were easily able to extend the advanced electrokinetic calculations with guidance. They also spotted typos in the equations of Ganguly et al.\cite{GangulyRoychowdhuryGupta2024}. The key challenge was that the output was mathematically dense, and it was quite challenging to parse the details of the calculations. For many tasks, we had to prompt and ask for the result to be made accessible, for instance, \textit{``The derivation should be accessible to a graduate student or a postdoctoral researcher. The governing equations and boundary conditions should be clear.''} and \textit{``This is too technical. Explain to me in 500-700 words on the overall potential of solving the problem using this approach''}. A second kind of prompt was needed simply to establish what had already been found, because the result was buried in the output that reported it: \textit{``Can you explain the status of the project in 500 words in accessible language''} and \textit{``The write up is quite confusing and dense. Before you do anything else, can you write the expression of the final velocity in terms $D_i$, $z_i$, $c_{i\infty}$ and their summations, and a function of $\kappa$? I am just thinking to understand the structure''}. We also had to simplify the answer in a presentable form by asking it to combine the result in a clean form, as the solver yielded a barely readable symbol choice. For instance, we asked \textit{``I also don't understand your variable naming as it will confuse a reader''} and then fixed the scheme ourselves, \textit{``maybe we separate with $\tilde\zeta$, $\tilde\zeta^2$, $\tilde\zeta^3$ terms, and then name them with supercript related to the $\tilde\zeta$ order, and the subscript w.r.t. the gradient term multiplying it ...''}, which is the naming used in eqn~\eqref{eq:dpanswer}.


\section{Bipolar membranes}
\label{sec:bpm}

\subsection{Background}

A bipolar membrane (BPM) consists of a cation-exchange layer (CEL) and an anion-exchange layer (AEL) joined at a junction\cite{Frilette1956}. Its $I$--$V$ curves show a rectifying response, which makes it look analogous to a semiconductor p-n junction\cite{Lovrecek1959,Shockley1949,Bassignana1983}. There is an electrolyte solution next to the open ends of both CEL and AEL. A dissociation catalyst is often placed at the junction\cite{Bui2024}. Under forward bias, the electrolyte's counter-ions are driven towards the junction between the layers. Under reverse bias, they are pulled away from the junction; the junction is depleted of mobile salt, and protons and hydroxide ions generated from water carry the current instead\cite{Parnamae2021,Bui2024}. Reverse bias is the mode applications use because it turns an applied voltage into a pH gradient. Some applications of BPM in reverse bias include acid and base recovery\cite{Wei2012,Zabolotskii2014}, CO$_2$ capture from air and seawater\cite{Bui2023}, water electrolysis\cite{Vermaas2018,Marin2023}, and energy storage\cite{Weng2018,Blommaert2021}, among others.

The reverse-bias curve has a recognisable shape\cite{Mafe1997,Wilhelm2002}. At low voltage, the response is ohmic, and the current is carried by salt co-ions leaking across the two layers. That leakage sets a limiting current $I_{\rm lim}$, i.e., a plateau in the $I$--$V$ curve reached once salt cannot reach the junction fast enough to hold the current\cite{Strathmann1997}. Past a threshold voltage, the current climbs again, and this overlimiting branch is where the device operates. Its slope is read as a water dissociation resistance, and a fourth region beyond it is attributed to the rate at which water itself reaches the junction\cite{Parnamae2021,Bui2024}.

\subsection{Literature}

The mechanisms offered for the overlimiting branch typically rely on a change in rate constants of the water-splitting reaction. For instance, one argument is Onsager's second Wien effect\cite{Onsager1934}, in which the large field in the depleted junction, reaching $10^{8}$~V~m$^{-1}$, raises the dissociation rate constant of water roughly exponentially\cite{Mafe1998}. The second argument is that water-splitting proceeds in two steps, in which a fixed charged group or an added catalyst accepts a proton from one water molecule, leaving OH$^-$ behind, and then donates that proton to another, making H$_3$O$^+$\cite{Simons1985,Zabolotskii1988}. The group returns to its original state at the end of each cycle, so it acts as a catalyst, and the rate then depends on the chemistry of the group rather than on the field alone. This mechanism is supported by the gains that follow from placing a catalyst at the junction\cite{McDonald2014,Oener2020,Chen2022,Chen2023}.

The transport of ions is treated separately. Continuum models built on the Nernst-Planck and Poisson equations\cite{Ramirez1992,Kovalchuk2006,Mareev2020,Bui2020,Bui2021} describe the junction either as a contact between the two layers or as a thin neutral film between them\cite{Strathmann1997}. Recent reports suggest that dilute-solution theory is likely inaccurate at the current densities industrial operation needs\cite{Bui2024}. Concentrated-solution theory is a promising alternative, but comes at the cost of more transport coefficients\cite{Bui2024}. Nonetheless, models still omit water transport, asymmetry between the layers, and the diffusion boundary layers outside the membrane\cite{Xu2002,Femmer2015,Parnamae2021,Bui2024}.

We got interested in the field because the junction inside bipolar membranes is much thinner than the cation and anion exchange layers. Therefore, it seemed like a perfect opportunity to do matching across the junction. While working on this problem, we came across the Parnamae et al.\cite{ParnamaePorada2023} model, which approached the problem through a similar idea. In this paper, water is held at chemical equilibrium everywhere, with no rate law for dissociation, and it still produces an overlimiting branch. If no kinetic step is present in the equations, something must be producing the current. In fact, the paper explicitly states that \textit{``our model that we used to obtain Figure 2 panels b and d does not include any description of the kinetic rate of water dissociation. Instead, we assume that the kinetics of the water dissociation reaction are infinitely fast. Our results indicate that we do not need to include a possible role of the electric field on the rate of water dissociation to simulate current-voltage curves that exhibit a clear overlimiting region with a current `takeoff'.''} However, the paper does not fully explain the mechanism, and it says so of its own analytical model, which \textit{``predicts well all the I-V curve regions where salt ions carry majority of the current: the forward bias region and the limiting current region, but not the overlimiting region where current is mainly carried by the H$^+$ and OH$^-$ ions''}. This was the motivation for our analysis.

\begin{figure}[t]
  \centering
  \includegraphics{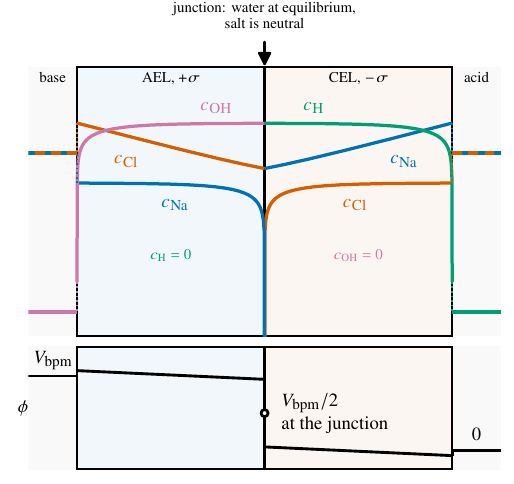}
  \caption{The bipolar membrane problem. A symmetric cell under reverse bias, drawn qualitatively. An anion-exchange layer of fixed charge $+\sigma$ and a cation-exchange layer of fixed charge $-\sigma$ meet at a junction, and each Donnan-equilibrates at its outer face against the same neutral salt bath, of salt concentration $c$ and proton concentration $c_w$. We assume that the proton is absent from the anion-exchange layer and the hydroxide from the cation-exchange layer. Only the acid half is solved; the base half is its mirror under $x\to-x$ with Na$^+$ for Cl$^-$ and H$^+$ for OH$^-$.}
  \label{fig:bpmsetup}
\end{figure}

\begin{figure*}[t]
  \centering
  \includegraphics[width=0.65\textwidth]{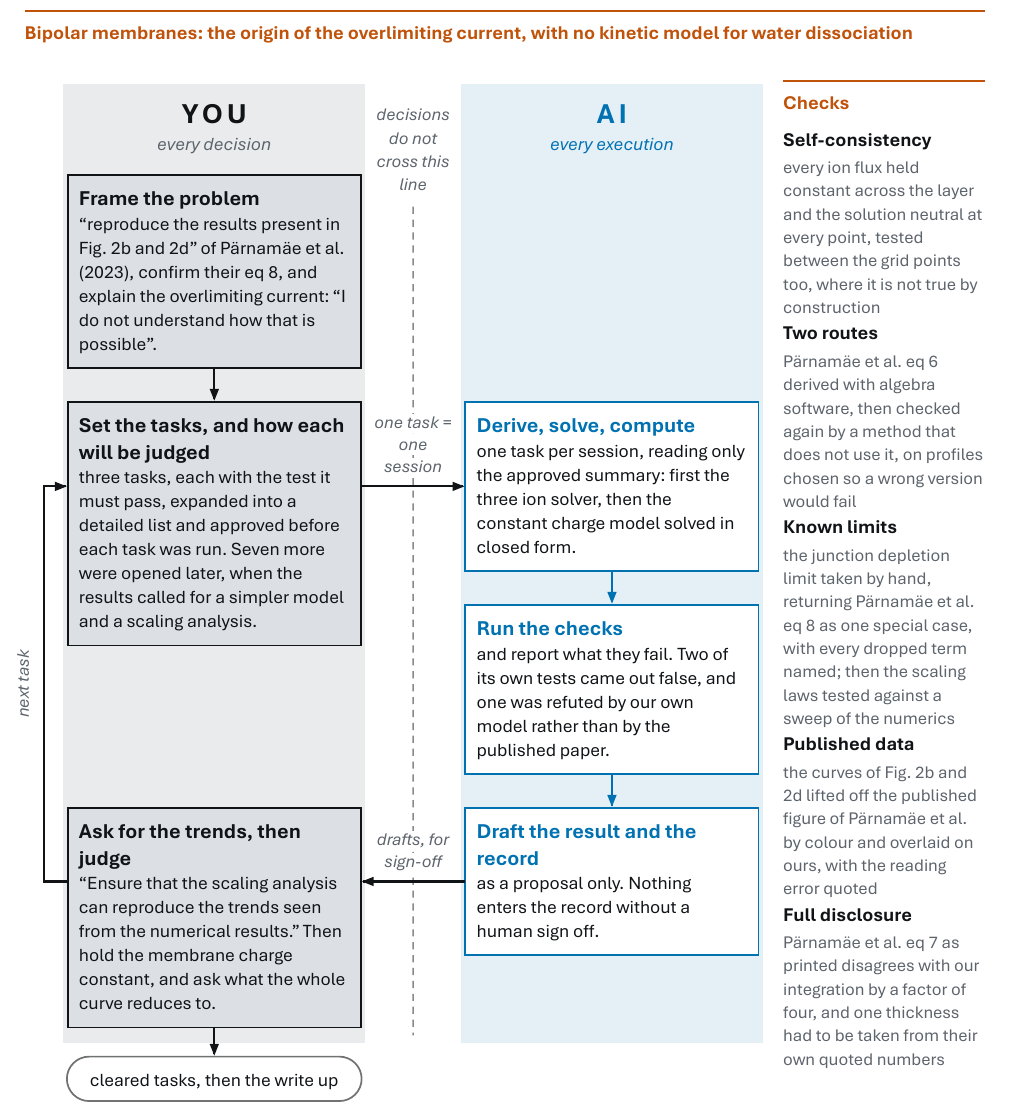}
  \caption{The framework of Fig.~\ref{fig:framework} as it was implemented for the bipolar membrane problem. Appendix~\ref{app:goals} reproduces the goals and tasks file that set them.}
  \label{fig:bpmcase}
\end{figure*}

\subsection{Problem setup and how AI tools were used}

A schematic of the problem is provided in Fig.~\ref{fig:bpmsetup}. Since we were inspired by the Parnamae et al. model\cite{ParnamaePorada2023}, we adapted our basic setup from their work, which assumes symmetry in the problem so that one has to solve only half of the geometry. We assume AEL and CEL carry the charge $\sigma$ and $-\sigma$, respectively. The acidic and basic reservoirs contain NaCl with concentration $c$ at the end, such that $c_{\rm Cl} = c_{\rm Na}=c$. The reservoirs are also pH neutral, such that H$^+$ and OH$^-$ are taken as $c_{\rm H} = c_{\rm OH} = c_w$. Hydroxide ions are assumed to be absent inside CEL such that $c_{\rm OH}=0$. Similarly, protons are assumed to be absent inside AEL such that $c_{\rm H}=0$. Diffusivities of ions $D_{\rm Na} = D_{\rm Cl}$ and $D_{\rm H} = D_{\rm OH}$ are also assumed to be equal. The voltage drop is also symmetric such that between the reservoir and the junction, both the layers experience a drop of $V_{\rm bpm}/2$. Finally, the junction is assumed to be at pH=7 and neutral so that $c_{\rm H} = c_{\rm OH} = c_w$ and $c_{\rm Na} = c_{\rm Cl} = c_j$. Due to the symmetry, the fluxes of Na$^+$ and Cl$^-$ are equal in magnitude and opposite. Just to reiterate, the objective was to understand the $I$--$V$ relationship from this model, preferably analytically. As shown in Fig.~\ref{fig:bpmiv} and detailed later, we are able to recover both the limiting and overlimiting current through a new analytical model based on a simple approximation.

The plan was set in advance before any calculation began; see Fig.~\ref{fig:bpmcase} and Appendix~\ref{app:goals} for the exact goals and tasks we gave to the AI tools, and Appendix~\ref{app:prompts} for representative prompts from the sessions themselves. The goal we stated at the top of that file was to \textit{``find a clear explanation, a scaling analysis and potentially an analytical result for overlimiting current in bipolar membranes where there is no kinetic model for water dissociation''}. Our first task was to ensure that we had the right set of equations, since they are not systematically laid out in Parnamae et al.\cite{ParnamaePorada2023}; we warned that \textit{``there are many inline equations, and they can be hard to follow since they aren't laid out in a pedagogical fashion. Be very careful when you read these.''} Next, we wanted to ensure that we could reproduce the results of Parnamae et al.\cite{ParnamaePorada2023}, and set the check for that task as \textit{``a one-to-one match with the results presented in Fig. 2b and 2d by digitizing the figure''}. We even explicitly stated that we did not understand how an overlimiting current could arise at all: \textit{``Investigate why one would get overlimiting current from these equations. I do not understand how that is possible, and if so, mathematically what is causing the overlimiting currents to occur and at what voltages, and why. This is the crucial piece.''} Finally, we wanted to look into scaling analysis and analytical approaches for the overlimiting current. These tasks were broken down further by AI tools into 16 tasks, which became twenty-two sessions and twenty-three task rows as seven more were opened along the way.

Later on, as we started understanding the physics better, we changed the problem. We fixed $\sigma$ instead of letting it be regulated, and we replaced the model's variable names with our own, \textit{``I labeled $\sigma$ as the membrane charge. And $c_w$ as $\sqrt{K_w}$ to indicate the neutral $c_{\rm H}$ concentration.''} We then enquired about the approximation that was finally used to create the analytical solution, first asking whether the proton could be dropped, \textit{``from the plot it seems clear to me that in the limiting and overlimiting current, we can simply forget about `H$^+$' ion and focus mostly on the Na$^+$ and Cl$^-$ ...''}, and then whether the co-ion could be, \textit{``Can I simply drop `Cl$^-$' from my equation and get to the overlimiting current branch?''} The exchange that followed was not useful, \textit{``Okay you are confusing me more''}, and what settled it was stating the calculation ourselves and asking only that it be executed: \textit{``create two models, one H$^+$ free for under and limiting, and one Cl$^-$ free for overlimiting, and then I will just want to overlay them on top of the numerical result''}. That is the calculation reported in the two subsections below.

\subsection{Governing equations and the solution procedure}

We only solve the acidic side of the system, which is the cation-exchange layer. We assume the junction to be at $x=0$ and the CEL/reservoir interface to be at $x=\ell$, where $\ell$ is the thickness of one layer.

Inside the layer, the three mobile species Na$^+$, Cl$^-$ and H$^+$ move by both diffusion and electromigration. Since there is no bulk reaction, each flux is independent of $x$. Writing $c_{\rm Na}$, $c_{\rm Cl}$ and $c_{\rm H}$ for the three concentrations, $J_{\rm Na}$, $J_{\rm Cl}$ and $J_{\rm H}$ for the corresponding fluxes, $\phi$ for the electric potential, $D=D_{\rm Na}=D_{\rm Cl}$ for the salt diffusivity, $D_{\rm H}$ for the proton diffusivity and $V_{\rm T}=RT/F$ for the thermal voltage, with $R$ the gas constant, $T$ the temperature and $F$ Faraday's constant, the Nernst-Planck fluxes are
\begin{equation}
\begin{aligned}
J_{\rm Na}&=-D\frac{{\rm d}c_{\rm Na}}{{\rm d}x}-\frac{Dc_{\rm Na}}{V_{\rm T}}\frac{{\rm d}\phi}{{\rm d}x},
\qquad
J_{\rm Cl}=-D\frac{{\rm d}c_{\rm Cl}}{{\rm d}x}+\frac{Dc_{\rm Cl}}{V_{\rm T}}\frac{{\rm d}\phi}{{\rm d}x},\\
J_{\rm H}&=-D_{\rm H}\frac{{\rm d}c_{\rm H}}{{\rm d}x}-\frac{D_{\rm H}c_{\rm H}}{V_{\rm T}}\frac{{\rm d}\phi}{{\rm d}x}.
\end{aligned}
\label{eq:bpmnp}
\end{equation}
The layer is locally electroneutral against its own fixed charge $-\sigma$,
\begin{equation}
c_{\rm Na}+c_{\rm H}-c_{\rm Cl}=\sigma ,
\label{eq:bpmen}
\end{equation}
and, as stated above, the symmetry of the cell makes the two salt fluxes equal and opposite,
\begin{equation}
J_{\rm Na}+J_{\rm Cl}=0 .
\label{eq:bpmsalt}
\end{equation}

\begin{figure*}[t]
  \centering
  \includegraphics{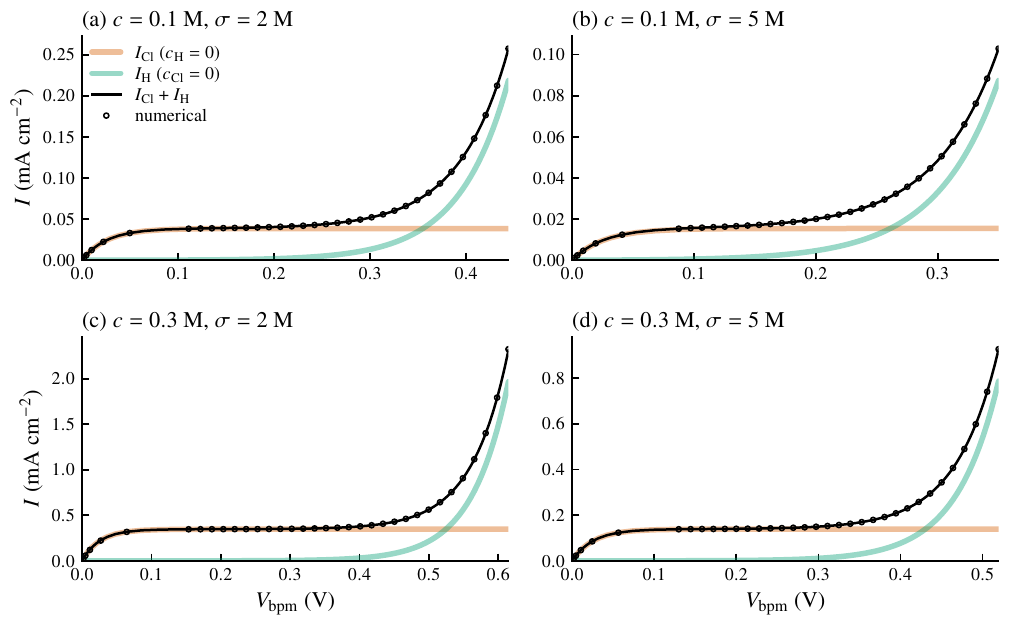}
  \caption{The $I$--$V$ curve at four combinations of $c$ and $\sigma$, with the analytical limits overlaid on the numerical solution. (a) $c=0.1$~M, $\sigma=2$~M; (b) $c=0.1$~M, $\sigma=5$~M; (c) $c=0.3$~M, $\sigma=2$~M; (d) $c=0.3$~M, $\sigma=5$~M, with $\ell=100~\mu$m, $c_w=10^{-7}$~M, $D=4\times10^{-11}$~m$^2$~s$^{-1}$, $D_{\rm H}/D=5$ and $T=298.15$~K throughout.}
  \label{fig:bpmiv}
\end{figure*}

Both ends of the layer are Donnan equilibria against a bath. At $x=\ell^-$ the bath is the acid reservoir, of salt concentration $c$ and proton concentration $c_w$, held at zero potential. Writing $\phi_D=\phi(\ell^-)$,
\begin{equation}
c_{\rm Na}(\ell^-)=c\,e^{-\phi_D/V_{\rm T}},\quad
c_{\rm Cl}(\ell^-)=c\,e^{+\phi_D/V_{\rm T}},\quad
c_{\rm H}(\ell^-)=c_w e^{-\phi_D/V_{\rm T}} .
\label{eq:bpmdonr}
\end{equation}
At $x=0^+$ the bath is the virtual solution occupying the junction plane, which is neutral and at pH 7 by the symmetry of the cell, so it carries salt concentration $c_{\rm j}$ and proton concentration $c_w$, and it sits at potential $V_{\rm bpm}/2$. Writing $\phi_0=\phi(0^+)$,
\begin{equation}
\begin{aligned}
c_{\rm Na}(0^+)&=c_{\rm j}e^{-(\phi_0-V_{\rm bpm}/2)/V_{\rm T}},\qquad
c_{\rm Cl}(0^+)=c_{\rm j}e^{+(\phi_0-V_{\rm bpm}/2)/V_{\rm T}},\\
c_{\rm H}(0^+)&=c_we^{-(\phi_0-V_{\rm bpm}/2)/V_{\rm T}} .
\end{aligned}
\label{eq:bpmdonj}
\end{equation}
The junction salt concentration $c_{\rm j}$ is unknown and is determined later. The current density is an output rather than a closure condition,
\begin{equation}
I=F\left(J_{\rm H}+J_{\rm Na}-J_{\rm Cl}\right) .
\label{eq:bpmcurrent}
\end{equation}
The problem is well-posed at a given $V_{\rm bpm}$. The equations were solved numerically by the shooting method. Fig.~\ref{fig:bpmiv} shows the computed $I$--$V$ curves.

\subsection{Key result}

We made analytical progress by combining two regimes. We first assume $c_{\rm H}\equiv0$, when eqn~\eqref{eq:bpmen} reduces to $c_{\rm Na}-c_{\rm Cl}=\sigma$. To make progress, we multiply the $c_{\rm Na}$ flux by $c_{\rm Cl}$, multiply the $c_{\rm Cl}$ flux by $c_{\rm Na}$, and add. The result becomes
\begin{equation}
c_{\rm Cl}J_{\rm Na}+c_{\rm Na}J_{\rm Cl}
=-D\left[c_{\rm Cl}\frac{{\rm d}c_{\rm Na}}{{\rm d}x}+c_{\rm Na}\frac{{\rm d}c_{\rm Cl}}{{\rm d}x}\right]
=-D\,\frac{{\rm d}}{{\rm d}x}\big(c_{\rm Na}c_{\rm Cl}\big) .
\label{eq:bpmcaseA0}
\end{equation}
Putting $J_{\rm Cl}=-J_{\rm Na}$ and $c_{\rm Cl}-c_{\rm Na}=-\sigma$ into the left-hand side of eqn~\eqref{eq:bpmcaseA0},
\begin{equation}
\frac{{\rm d}}{{\rm d}x}\big(c_{\rm Na}c_{\rm Cl}\big)=\frac{\sigma J_{\rm Na}}{D} ,
\label{eq:bpmcaseA1}
\end{equation}
whose right-hand side is a constant, so $c_{\rm Na}c_{\rm Cl}$ is linear in $x$.

At the reservoir face, eqn~\eqref{eq:bpmdonr} gives $c_{\rm Na}=ce^{-\phi_D/V_{\rm T}}$ and $c_{\rm Cl}=ce^{+\phi_D/V_{\rm T}}$, so their product is $c^2$; at the junction face, eqn~\eqref{eq:bpmdonj} gives $c_{\rm j}^2$ in the same way. Integrating eqn~\eqref{eq:bpmcaseA1} across the layer therefore gives $c^2-c_{\rm j}^2=\sigma J_{\rm Na}\ell/D$ directly, and since $J_{\rm H}=0$ and $J_{\rm Cl}=-J_{\rm Na}$ make eqn~\eqref{eq:bpmcurrent} read $I=2FJ_{\rm Na}$,
\begin{equation}
I_{\rm Cl}=\frac{2FD}{\ell}\,\frac{c^2-c_{\rm j}^2}{\sigma},
\qquad
I_{\rm lim}=\frac{2FDc^2}{\sigma\ell} ,
\label{eq:bpmIA}
\end{equation}
where $I_{\rm Cl}$ is the current when $c_{\rm H}\equiv0$, and $I_{\rm lim}$ is the limiting current. It is easy to see that the limiting current is obtained when $c_{\rm j}\to0$, or at a junction where salt vanishes.

Equation~\eqref{eq:bpmIA} provides the limiting current, but does not provide an $I$--$V$ relationship. Subtracting the two salt fluxes of eqn~\eqref{eq:bpmnp}, and using the fact that $c_{\rm Na}-c_{\rm Cl}=\sigma$ is a constant, we obtain
\begin{equation}
2J_{\rm Na}=J_{\rm Na}-J_{\rm Cl}
=-\frac{D}{V_{\rm T}}\big(c_{\rm Na}+c_{\rm Cl}\big)\frac{{\rm d}\phi}{{\rm d}x} ,
\label{eq:bpmcaseA2}
\end{equation}
which is exact. Solving $c_{\rm Na}c_{\rm Cl}=c^2$ with $c_{\rm Na}-c_{\rm Cl}=\sigma$ gives $c_{\rm Cl}(\ell^-)\simeq c^2/\sigma$, so $c_{\rm Cl}/\sigma$ is of order $(c/\sigma)^2$ and $c_{\rm Na}+c_{\rm Cl}\simeq\sigma$ throughout the layer. We can then write $\big(c_{\rm Na}+c_{\rm Cl}\big)\simeq\sigma$, and eqn~\eqref{eq:bpmcaseA2} has a constant right-hand side, so the field is uniform, and substituting $J_{\rm Na}$ from eqn~\eqref{eq:bpmIA} gives the whole drop across the layer,
\begin{equation}
\frac{\ell}{V_{\rm T}}\left|\frac{{\rm d}\phi}{{\rm d}x}\right|
=\frac{2\left(c^2-c_{\rm j}^2\right)}{\sigma^2}\le\frac{2c^2}{\sigma^2}\ll1 .
\label{eq:bpmcaseA3}
\end{equation}
There is essentially no potential drop across the cation-exchange layer. Therefore, one can simply write $c_j$ using a Donnan jump of $V_{\rm bpm}/2$ from the junction to the reservoir, or
\begin{equation}
c_{\rm j}=c\,e^{-V_{\rm bpm}/2V_{\rm T}} ,
\label{eq:bpmcj}
\end{equation}
and eqn~\eqref{eq:bpmIA} becomes
\begin{equation}
I_{\rm Cl}=I_{\rm lim}\left(1-e^{-V_{\rm bpm}/V_{\rm T}}\right).
\label{eq:bpmIAthin}
\end{equation}
Equation~\eqref{eq:bpmIAthin} is the textbook limiting-current curve.

Next, we move to the overlimiting branch. Here we assume $c_{\rm Cl}\equiv0$. Then $J_{\rm Cl}=0$, and eqn~\eqref{eq:bpmsalt} forces $J_{\rm Na}=0$ as well. Therefore, the whole current is carried by the protons, which we define as $I_{\rm H}=FJ_{\rm H}$. Three consequences follow from the assumptions.

First, $J_{\rm Na}=0$ puts the sodium in equilibrium. Setting the first of eqn~\eqref{eq:bpmnp} to zero,
\begin{equation}
\frac{{\rm d}c_{\rm Na}}{{\rm d}x}=-\frac{c_{\rm Na}}{V_{\rm T}}\frac{{\rm d}\phi}{{\rm d}x} .
\label{eq:bpmcaseB0}
\end{equation}
Second, electroneutrality eqn~\eqref{eq:bpmen} is now $c_{\rm Na}+c_{\rm H}=\sigma$ with $\sigma$ constant, so ${\rm d}c_{\rm H}/{\rm d}x=-{\rm d}c_{\rm Na}/{\rm d}x$. Substituting that and eqn~\eqref{eq:bpmcaseB0} into the proton flux of eqn~\eqref{eq:bpmnp}, the two concentrations combine into their sum,
\begin{equation}
J_{\rm H}=-D_{\rm H}\left[\frac{{\rm d}c_{\rm H}}{{\rm d}x}+\frac{c_{\rm H}}{V_{\rm T}}\frac{{\rm d}\phi}{{\rm d}x}\right]
=-\frac{D_{\rm H}}{V_{\rm T}}\big(c_{\rm Na}+c_{\rm H}\big)\frac{{\rm d}\phi}{{\rm d}x}
=-\frac{D_{\rm H}\sigma}{V_{\rm T}}\frac{{\rm d}\phi}{{\rm d}x} .
\label{eq:bpmcaseB1}
\end{equation}
Since $J_{\rm H}$ does not depend on $x$, neither does ${\rm d}\phi/{\rm d}x$: the field is uniform. Integrating, we obtain
\begin{equation}
\phi_0-\phi_D=\frac{\ell J_{\rm H}V_{\rm T}}{D_{\rm H}\sigma} .
\label{eq:bpmcaseB1b}
\end{equation}
Third, putting eqn~\eqref{eq:bpmcaseB1} back into eqn~\eqref{eq:bpmcaseB0} gives ${\rm d}c_{\rm Na}/{\rm d}x=[J_{\rm H}/(D_{\rm H}\sigma)]c_{\rm Na}$, and integrating we obtain
\begin{equation}
J_{\rm H}=\frac{D_{\rm H}\sigma}{\ell}\ln\frac{c_{\rm Na}(\ell^-)}{c_{\rm Na}(0^+)} .
\label{eq:bpmcaseB2}
\end{equation}

At the reservoir face, eqn~\eqref{eq:bpmdonr} gives $c_{\rm Na}$ and $c_{\rm H}$ as $c$ and $c_w$ multiplied by the same factor $e^{-\phi_D/V_{\rm T}}$, and $c_{\rm Na}+c_{\rm H}=\sigma$ fixes that factor at $\sigma/(c+c_w)$; the junction face works identically through eqn~\eqref{eq:bpmdonj}, with $c_{\rm j}$ in place of $c$. Hence
\begin{equation}
c_{\rm Na}(0^+)=\frac{\sigma c_{\rm j}}{c_{\rm j}+c_w},
\qquad
c_{\rm Na}(\ell^-)=\frac{\sigma c}{c+c_w} .
\label{eq:bpmcaseB3}
\end{equation}
Since $J_{\rm Na}=0$, the sodium at the junction is in equilibrium with the reservoir, or $c_{\rm j}=c\,e^{-V_{\rm bpm}/2V_{\rm T}}$. Substituting into eqns~\eqref{eq:bpmcaseB2} and \eqref{eq:bpmcaseB3} reveals
\begin{equation}
I_{\rm H}=\frac{FD_{\rm H}\sigma}{\ell}\ln\!\left[\frac{c+c_we^{V_{\rm bpm}/2V_{\rm T}}}{c+c_w}\right].
\label{eq:bpmIB}
\end{equation}

The total current can be calculated as the plain sum of the two limits,
\begin{equation}
\begin{aligned}
I\simeq I_{\rm Cl}+I_{\rm H}
={}&\frac{2FDc^2}{\sigma\ell}\left(1-e^{-V_{\rm bpm}/V_{\rm T}}\right)\\
&+\frac{FD_{\rm H}\sigma}{\ell}\ln\!\left[\frac{c+c_we^{V_{\rm bpm}/2V_{\rm T}}}{c+c_w}\right].
\end{aligned}
\label{eq:bpmtotal}
\end{equation}

Comparison of eqn~\eqref{eq:bpmtotal} for a variety of $\sigma$ and $c$ values is shown in Fig.~\ref{fig:bpmiv}. As is observed from the figure, we can capture all essential trends based on our analysis.

One interesting quantity of interest is the take-off voltage. The take-off voltage follows from eqn~\eqref{eq:bpmtotal}. Once $c_we^{V_{\rm bpm}/2V_{\rm T}}\gg c$ the logarithm is linear in $V_{\rm bpm}$, so the overlimiting branch approaches a straight line whose intercept on the voltage axis is
\begin{equation}
V_0=2V_{\rm T}\ln\frac{c+c_w}{c_w}\simeq2V_{\rm T}\ln\frac{c}{c_w} .
\label{eq:bpmV0}
\end{equation}
Its physical meaning follows from eqn~\eqref{eq:bpmcj}: $V_0$ is the voltage at which the junction salt $c_{\rm j}=ce^{-V_{\rm bpm}/2V_{\rm T}}$ has been depleted to the water-ion level $c_w$.

\subsection{Experience with AI tools}

We observed that AI tools were able to easily reproduce the trends of Parnamae et al.\cite{ParnamaePorada2023}. The tools were also able to help confirm that a constant $\sigma$ value also yields overlimiting current. However, AI tools were less useful in physical interpretation and scaling analysis. While their math was accurate, their physical explanation was difficult to follow. We provided several inputs. On the scaling analysis, we could barely understand it, \textit{``I am confused with the scaling analysis that the session was pushing. I want to understand how to put together a coherent scaling analysis''} and, later, \textit{``To be clear, is there a pure explicit solution for all the variables? These scaling analysis do not seem to make much sense to me''}. On the physical explanation, the difficulty was length rather than error, \textit{``You write very long responses that end up confusing more than helping. Physically, help me understand why I should assume $\Delta\mu_{\rm Na^+}$ to be zero''}. We had to enquire about an easier explanation such as \textit{``Give me the response a technical level accessible to a grad student, not too generic. Give key math results and explain them physically''} and \textit{``walk me through in a detailed step by step manner ... Imagine no prior knowledge so explain scaling analysis clearly step by step so that I can fully understand the different results that you are providing''}.


\section{More details about AI usage and thoughts on graduate education}
\label{sec:education}

Overall, the project took about 68 million tokens (approximately \$2100 at Claude Opus 5 API rates, counting the cached context re-read on every request; the authors were on a subscription plan and were not billed per token) and about 75 hours of active session time, spread over two weeks of calendar time. The split across the three parts was about 19 million tokens, \$670 and 26 hours for the diffusiophoresis analysis; about 27 million tokens, \$730 and 27 hours for the bipolar membrane analysis; and about 22 million tokens, \$740 and 22 hours for the write-up, the figures and the record-keeping. We emphasize that this speed is reported for transparency but should not be used as a benchmark. We had thought about both problems, had extensive experience in the areas, and had a clear plan formulated before starting. Without this experience, it would have been challenging to guide the AI tools and parse apart their output.

Separately, we would like to note that AI's usage in shaping the narrative and writing was minimal. We did use it to organize references, create figures, and write out the equations. However, we have found that AI tools cannot help one write and can, in fact, push a narrative or framing that is not accurate. They can be helpful for grammar checking, spell-checking, and fact-checking.

We have provided a range of files along with the manuscript as Supplementary Information, described in Section~\ref{sec:si}. We invite the reader to look at them.

Given that we were able to solve a PhD-level problem with AI assistance, we offer some thoughts on graduate training. These tools are remarkably capable, and a student can extract a great deal of value from them. That is precisely why we believe training should not be compromised. The skills the tools now perform quickly, such as carrying an expansion through by hand, setting up and debugging a numerical solution, and reading a paper closely enough to find its errors, are the same skills a researcher needs in order to judge whether the output is right. A student who has never developed them will find it difficult to differentiate between a correct derivation and a plausible one. In addition, the repetition of those skills is precisely what builds intuition to make sense of the math. We do not think AI can replace that, even if it can solve things much faster.

One practical option is to complete a couple of projects during the PhD without the assistance of AI, so that the basic skill sets of algebra, numerical computation, and careful reading are acquired first. These could be reproducing a known result in the field, or something that your research group has recently been working on. Alongside that, we suggest reading the literature directly and forming your own sense of the field rather than accepting a summary of it using AI. As the two sections above describe, the difficulty in this project was rarely the mathematics itself; it was following the physical argument and knowing which parts to trust. The risk is certainly cognitive offloading, where an answer is obtained but the learning that comes along with it is lost.

We believe AI will be used heavily in the research of the future. However, a researcher's expertise to judge and guide a scientific project is likely going to be even more valuable than in pre-AI times. The tool is only as good as the user guiding it. For transparency, we have provided representative prompts in Appendix~\ref{app:prompts} for an interested reader about how we guided it.


\section{Conclusion}
\label{sec:conclusion}

For diffusiophoresis, we obtained the velocity of a charged sphere in a mixture of an arbitrary number of electrolytes at arbitrary Debye length, in the Debye--H\"uckel limit and to $O(\tilde\zeta^3)$. The result separates into seven universal functions of $\kappa$ multiplying combinations of the valence-weighted moments of the far field, so that the double-layer physics is computed once and reused for any mixture. It reduces to Henry's function and to the chemiphoretic coefficient of Keh and Wei\cite{KehWei2000} for a binary salt, and to eqn~(26) of Gupta et al.\cite{GuptaRallabandiStone2019} in the thin-layer limit. Some potential extensions in the future include relaxing the small P\'eclet number limit and focusing on shape effects\cite{GangulyGupta2026}.

For bipolar membranes, we obtained a closed-form $I$--$V$ relationship in reverse bias with water held at equilibrium and no kinetic rate law for its dissociation. We derived eqn~\eqref{eq:bpmtotal}, which can capture the numerical solution of the full three-ion problem. The overlimiting current therefore requires neither a field-enhanced dissociation rate nor a catalytic step; it is the proton current through a junction whose salt has been depleted below the water-ion concentration, with a take-off voltage, eqn~\eqref{eq:bpmV0}. Future work could include solving asymmetric membranes, including catalytic activity, and non-ideal transport equations.


\section{Supplementary files}
\label{sec:si}

We provide four things with the article: the rules the AI tool worked under, the figure rules, the command that closed every session, and the scripts that produce what the article shows.

\begin{itemize}\setlength{\itemsep}{3pt}
\item \texttt{CLAUDE.md}, the file of instructions read at the start of every session, where the honesty rules of Section~\ref{sec:framework} are written.
\item \texttt{PlotRules.md} and \texttt{shared/}, the plotting rules and the style files that every figure script is held to.
\item \texttt{wrap.md}, the command that writes the session record, updates the task status and the summary, and proposes correction entries for approval.
\item the scripts, one folder per part of the project. These are a selection rather than all of them: of the 119 scripts written, we provide the 38 that carry a reader from the starting papers to the two results and the eight figures, and leave out the probes, the digitisations of published figures, the superseded models and the figures that never reached the article. Every number in every figure comes from one of the 38. The write-up folder also holds four scripts that check the article rather than the physics: one compares the master text against the typeset copy word by word, one tests every prompt quoted here against the goals files and the session transcripts, one measures the usage reported in Section~\ref{sec:education}, and one traces numbers printed in the figures back to the file they came from. A \texttt{README.md} in the bundle names every script it holds and says what each one does.
\end{itemize}


\section*{Acknowledgements}

A.G. thanks the NSF CAREER program (CBET-2238412) and the Air Force Office of Scientific Research (Grant No. FA9550-25-1-0176, Young Investigator Program Award) for financial support.

\bibliography{refs}
\bibliographystyle{unsrtnat}


\appendix

\section{Goals and tasks, as given to the AI}
\label{app:goals}

The two files below are the complete goals-and-tasks files for the two problems. Each was written before any calculation began and was the first thing the AI tool was given, and each task carries the check it had to pass. They are reproduced word for word, spelling and punctuation as typed; the only change is that the line breaks have been reflowed to the column width.

\subsection{Diffusiophoresis}

\begin{itemize}\setlength{\itemsep}{3pt}
\item \textit{Goal: Find diffusiophoretic mobility of a spherical particle in a mixture of electrolytes for arbitrary Debye lengths}
\item \textit{Task-1: Go to 1-diffusiophoresis folder and read the document at relevant publications/\allowbreak{}2024\_\allowbreak{}GangulyRoychowdhuryGupta\_\allowbreak{}JFM.pdf. Specifically, I want you to go through section 3.3 carefully and understand the governing equations and boundary conditions. Then, I want you to reproduce the derivation and the numerical results in Fig. 3.}
\item \textit{Check for task 1: A one-to-one match with the results presented in Fig. 3. Also, ensure that all other equations are recovered correctly.}
\item \textit{Deliverable: Create a figure matching the results presented in the paper by digitizing the Figure in the paper, and overlaying the numerical results that you will derive. In addition, create a LaTex file in the format of Soft Matter, and write down the full derivation for me to check, while also including the figure.}
\item \textit{Task-2: I want to rewrite the governing equations and boundary conditions for an arbitrary number of ions with arbitrary diffusivities and valences. Confirm if it is possible to solve the problem only by solving ``salt'', ``charge'', and ``ionic strength'', and maybe one or two more variables. Discuss with me if clarifications are needed.}
\item \textit{Check for task-2: The equations should relax to the binary electrolyte scenario.}
\item \textit{Deliverable: Include the equations in the LaTex file from Task-1.}
\item \textit{Task-3: Perform the perturbation analysis and collect the governing and boundary conditions. You can go up to the order of zeta potential\textasciicircum{}3 in perturbation analysis.}
\item \textit{Check for task-3: Ensure all the equations and boundary conditions are consistent with the conservation equations. Check if the problem is well-posed.}
\item \textit{Deliverable: Include the equations in the LaTex file.}
\item \textit{Task-4: Solve the equations, potentially both numerically and analytically, and arrive at the equation.}
\item \textit{Check for task-4: Reduce to binary electrolyte result. Reduce to Eq. 26 in 2019\_\allowbreak{}GuptaRallabandiStone\_\allowbreak{}PRFluids.pdf inside the ``relevant publications'' folder. Note that Eq. 26 in that paper is for diffusioosmosis and is thus going to have a negative sign and should be consistent with the result at the thin double layer limit only.}
\item \textit{Deliverable: Include the equations in the LaTex files. Make sure verifications are also detailed.}
\end{itemize}

\subsection{Bipolar membranes}

\begin{itemize}\setlength{\itemsep}{3pt}
\item \textit{Goal: Find a clear explanation, a scaling analysis and potentially an analytical result for overlimiting current in bipolar membranes where there is no kinetic model for water dissociation}
\item \textit{Task-1: Go to the 2-bipolar-membranes folder and read the document in relevant publications/\allowbreak{}2023\_\allowbreak{}ParnamaePorada\_\allowbreak{}EST. Specifically, I want you to go through the section ``Theory of Ion Transport in Bipolar Membranes'' carefully and understand the governing equations and boundary conditions. There are many inline equations, and they can be hard to follow since they aren't laid out in a pedagogical fashion. Be very careful when you read these. I want to make sure that those are consistently laid out. Then, I want you to reproduce the results present in Fig. 2b and 2d by developing a numerical solver.}
\item \textit{Check for task 1: A one-to-one match with the results presented in Fig. 2b and 2d by digitizing the figure.}
\item \textit{Task-2: I want to investigate the analytical result further and understand the details. Reproduce it and confirm equation 8.}
\item \textit{Check for task-2: The equations should reduce to equation 8 in the manuscript.}
\item \textit{Task-3: Investigate why one would get overlimiting current from these equations. I do not understand how that is possible, and if so, mathematically what is causing the overlimiting currents to occur and at what voltages, and why. This is the crucial piece. I want you to carefully go through this analysis be re-dimenionsalizing the equations with variables such as the length of the membrane, concentration of the electrolyte, diffusivity of the ions, etc., and then coming up with a scaling analysis to explain why overlimiting current is occurring. If there is a potential for analytical progress for overlimiting current, explain what that would look like, and then I will tell if it makes sense to proceed with it.}
\item \textit{Check for task-3: Ensure that the scaling analysis can reproduce the trends seen from the numerical results.}
\item \textit{Deliverable: Create a figure matching the results presented in the paper by digitizing the Figure in the paper and overlaying the numerical results that you will create. In addition, create a LaTeX file in the format of Soft Matter, and write down a fully detailed setup of governing equations and boundary conditions for me to check, including the numerical protocol. Then, focus on the analytical result for the limiting current. Finally, focus on the overlimiting current scaling analysis and analytical result, if possible.}
\end{itemize}

\section{Representative prompts from the sessions}
\label{app:prompts}

The prompts below were typed during the sessions, as distinct from the plan in Appendix~\ref{app:goals}, and every one is reproduced verbatim, spelling and punctuation as typed; where a prompt is shortened, the elision is marked ` ...'. The prompts quoted in Sections~\ref{sec:diffusiophoresis} and \ref{sec:bpm} are a subset of the same record and are not all repeated here. Each is labelled DP for the diffusiophoresis problem and BPM for the bipolar membrane problem.

\subsection{One task per session, and the running record}

\begin{itemize}\setlength{\itemsep}{2pt}
\item BPM. \textit{``Read goals and tasks.txt file. Based on reading the file, create tasks.md, checks.md and corrections.md files. Clarify anything if needed.''}
\item DP. \textit{``Read the tasks.md, checks.md, summary.md and corrections.md files, and then perform task 3''}
\item BPM. \textit{``Read the tasks.md, checks.md, summary.md and corrections.md files, and then performed task 12''}
\item BPM. \textit{``Okay. Push wrap but briefly note that this discussion took place. I would like to return to this physical picture after Task-14.''}
\item BPM. \textit{``record the summary and the correction I noticed into the files by pushing wrap. I want to get deeper into the scaling in the next session''}
\item DP. \textit{``Read the tasks.md, checks.md, summary.md and corrections.md files. Run task 19. My understanding was that this was already done in task 18. Confirm if so and then we can finish this session''}
\end{itemize}

\subsection{Calling the verdict, and setting a task aside}

\begin{itemize}\setlength{\itemsep}{2pt}
\item DP. \textit{``This is fine. I would call this a pass, just note that the outer BC of 1+100/kappa causes a slightl discrepancy, but that is the extend of it. Once you do that, you can run wrap to close out the session''}
\item DP. \textit{`I cannot fully review the draft since it is quite long, so for now, write that it is a ``conditional pass'' that I will return to later. After that, run the wrap'}
\item DP. \textit{``Read the tasks.md, checks.md, summary.md and corrections.md files. Tasks 1-8 are done. Skip task 9 entirely. Record that it is not required because all the other checks have satisfied me, so I do not need to see the detailed derivation. Record this change, pass all the checks.''}
\item DP. \textit{``Read the tasks.md, checks.md, summary.md and corrections.md files. Instead of doing task 16, let us table it as it will be hard to read, so record that I do not want to to do this now and push wrap about task 16 with this message.''}
\item BPM. \textit{``Read the tasks.md, checks.md, summary.md and corrections.md files, and then read task 14. Before you go deep into this, my understanding is that this was already done in task 13. When I say I want to find closed form solutions, I am looking for concentration profiles, current and voltage takeoff values, which seems to have been found in task 13. First, confirm if true. If so, I do not think task 14 is required.''}
\item BPM. \textit{``Yes, that is fine. Let us mark this dead. But can you overlay whatever analytical results you already have on top of your own numerical results and create a figure so that I am confident that the results are looking okay? I know they will disagree with the numerical ones, but I just want to see how they look visually''}
\end{itemize}

\subsection{Scoping and steering the calculation}

\begin{itemize}\setlength{\itemsep}{2pt}
\item DP. \textit{``Note that the expansion of GRS may not be correct but the full analytical solution should be available to cross-check accurately''}
\item BPM. \textit{``Can you plot concentration profiles of C\_Na, C\_Cl, C\_H for me in underlimiting, limiting, and overlimiting currents? Perhaps take two representative voltages each.''}
\item BPM. \textit{`So, use lowercase ``c'' instead of ``C''. I do not want anything eliminated. All J\_H, J\_Na, J\_Cl are indeed constants. First, just list out the equations on how you would write them here so that I can double check everything. Do not eliminate yet.'}
\item BPM. \textit{`read the summary of P2S20 and use those conventions to create a write up of the key equations listed there; simply state them. Then solve these numerically, and then the two limits anylitically with the one limit with c\_H=0 and the other being c\_Cl=0, and the idea is to superpose the curves and show it matches numerically for a range of ``c'' values. You can read prior session files to know the values of other parameters'}
\end{itemize}

\subsection{Asking for the physics, and for plain language}

\begin{itemize}\setlength{\itemsep}{2pt}
\item BPM. \textit{``Okay all of this looks good and I could follow along. Can you now look at the two expressions for the two two-ion problems and physicall explain in simplest possible terms what is going on?''}
\item BPM. \textit{``Okay I understand this physically. Can you explain the linearity of the overlimiting current...I thought it would be exponential but seems that is incorrect.''}
\item BPM. \textit{``Can you explain in simple words this whole idea of H+ boundary layer and overlimiting current. Why would there be an overlimiting current at all, and what is causing it in this model''}
\item BPM. \textit{``Okay - I like this but walk me through in a detailed step by step manner. Some of the equations aren't clarified fully. I want to understand all the different limits. Imagine no prior knowledge so explain scaling analysis clearly step by step so that I can fully understand the different results that you are providing''}
\item DP. \textit{``Can you explain the status of the project in 500 words in accessible language''}
\item DP. \textit{`Could you physically explain with ``A'' ends up influencing u\_2 and u\_3 at finite kappa*a ...it is not as intuitive of a result'}
\end{itemize}

\subsection{Guiding and pushing back, if needed}

\begin{itemize}\setlength{\itemsep}{2pt}
\item BPM. \textit{``Wait -- why are we solving the jump at the interface? wouldn't that be calculated from the Donnan equilibrium?''}
\item BPM. \textit{``From what I understand from the derivation, charge regulation is not necessarily required for the overlimiting current. First, is that correct? I am trying to simplify the set up to get a cleaner result''}
\item BPM. \textit{``Can you explain the well-posedness. J\_Na, J\_Cl, J\_H are three uknowns, and I have 6 boundary conditions for concentrations, with 3 equations. So that closes the loop. On top, I do not know c\_j and phi\_D, phi, and I do have electroneutrality and J\_Na + J\_Cl=0 , so I feel like I am one equation short. Can you help explain how this is well-posed?''}
\item BPM. \textit{``Please explain the statement $\hat{V}/2 = \delta \mu_H/(R*T)$ ...it doesn't seem to make sense to me''}
\item BPM. \textit{``You write very long responses that end up confusing more than helping. Physically, help me understand why I should assume Delta mu Na\textasciicircum{}+ to be zero. If it is in equilibrium, then it shouldn't be giving flux...My guess is that it is a first order approximation, but you do count it to give flux...I just dont see the consistency in these statements''}
\item BPM. \textit{``To be clear, is there a pure explicit solution for all the variables? These scaling analysis do not seem to make much sense to me''}
\item BPM. \textit{``Okay you are confusing me more.''}
\item BPM. \textit{``I uploaded the 1999 paper in relevant literature folder but do not believe it is the same scope; we kind of need the 3 ion variation to do it''}
\item DP. \textit{`What is the wrong ``scaling factor''? Can you expand on that?'}
\item DP. \textit{``It appears to me that this confusion is coming because Keh and Wei non-dimensionalized their velocity differently than the result in Ganguly et al JFM, and when the authors try to reconcile those, they have to apply those factors to bring them to partiy. Confirm if so.''}
\item DP. \textit{``How is the consistent with KW which had -21/2 kappa for zeta\textasciicircum{}2 from what I can recall?''}
\item DP. \textit{``Why do we need closed-form results? If it is such a function of kappa, it might be easier to plot numerically anyways?''}
\item DP. \textit{``I also don't understand your variable naming as it will confuse a reader''}
\end{itemize}

\end{document}